\documentclass{moriond}

\usepackage{amsmath,amssymb,color,slashed,graphicx}
\def\be{\begin{equation}}
\def\ee{\end{equation}}
\def\bea{\begin{eqnarray}}
\def\eea{\end{eqnarray}}

\newcommand{\Photo} 

\begin{document}
\vspace*{4cm}
\title{PROBING RADIATIVE NEUTRINO MASS MODELS WITH DILEPTON EVENTS AT THE LHC}

\author{ Chahrazed Guella, Dounia Cherigui}
\address{Facult\'e de Physique, d\'{e}partement de G\'enie Physique, USTO-MB, BP1505 El M'Naouar, Oran, Alg\'erie.}
\author{Amine Ahriche}
\address{Laboratory of Mathematical and Sub-Atomic Physics (LPMPS), University of Constantine I, DZ-25000 Constantine, Algeria}
\author{Salah Nasri}
\address{Department of Physics, United Arab Emirates University, POB 17551, Al-Ain, UAE.}
\author{Rachik Soualah}
\address{Department of Applied Physics and Astronomy, University of Sharjah, P.O. Box 27272 Sharjah, UAE}

\maketitle\abstracts{
In this work we prob a class of neutrino mass models at both Large Hadron Collider (LHC) energies 8 TeV and 14 TeV. The focus will be on the new introduced interaction terms between a singlet charged scalar, $S^{\pm}$, and leptons leading to different final states $pp\rightarrow\ell_{\alpha}^{\pm}\ell_{\beta}^{\mp}+ \slashed{E} $ with $\ell_{\alpha}\ell_{\beta}=ee,e\mu,\mu\mu$ that implies lepton flavor violation (LFV). An accurate cut on the $M_{T2}$ eventvariable is found to be crucial for an effective suppression of the large Standard Model background. The obtained results can be translated into a possible detectability of the charged scalars effect.}
 
\section{Introduction}

Nowadays one of the striking questions that remains still unexplained by the Standard Model is the smallness of the neutrino mass. Several models Beyond the SM could provide an explanation of the non-zero mass of neutrinos. Among these models, we focus on the radiative neutrino mass models which are the simplest way to generate a small mass for neutrinos at the loop level. To tackle this, the SM is extended with new interaction terms yielded by extra implemented scalars and fermions singlets and/or doublets through one loop, two loops, or three loops.
 

\section{Radiative neutrino mass models and charged scalars}
We introduce a class of SM extensions at the loop level with two electrically charged singlet fields, $S^{\pm}$, that transforms under the SM gauge group $SU(3)\times SU(2)_{L}\times U(1)_{Y}$ as $S^{+}\sim(1,1,2)$. The different Lagrangian interaction terms  that are described in Ref. \cite{guella} can be summarised as follows:

\begin{eqnarray}
\mathcal{L} \supset\{f_{\alpha\beta}\overline{L_{\alpha}^{c}}L_{\beta
}S^{+}+\mathrm{H.c.}\}-M_{S}^{2}S^{+}S^{-}-V(H,S,\phi_{i}),\label{L}\\
V(H,S,\phi_{i})  \supset\lambda_{HS}\left\vert H\right\vert ^{2}\left\vert
S\right\vert ^{2}%
\end{eqnarray}

where $L_{\alpha}=(\nu_{\alpha L},\ell_{\alpha L})^{T}$, $\ell_{\alpha R}$ is
the charged lepton singlet, $f_{\alpha\beta}$ are the Yukawa couplings, and $c$ indicates the charge conjugation operation. The SM Higgs field doublet is denoted by $H$ and any additional scalar representation(s) is represented by $\phi_{i}$. In this kind of model(s), the parameter sets must fulfill the LFV constraints as well as other constraints studied in \cite{Adam:2013mnn}.
  
  
\subsection{Production of the charged scalar $S^{\pm}$}\label{subsec:prod}

The charged scalars $S^{\pm}$ are produced in pairs in the proton-proton collisions via the Drell-Yan (DY) s-channel processes as:

\begin{equation}
q\overline{q}\rightarrow\gamma/Z/h\rightarrow S^{+}S^{-},~gg\rightarrow
h\rightarrow S^{+}S^{-},
\end{equation}

\begin{figure}[h]
\begin{centering}
\includegraphics[width=0.5\textwidth,height=4.5cm]{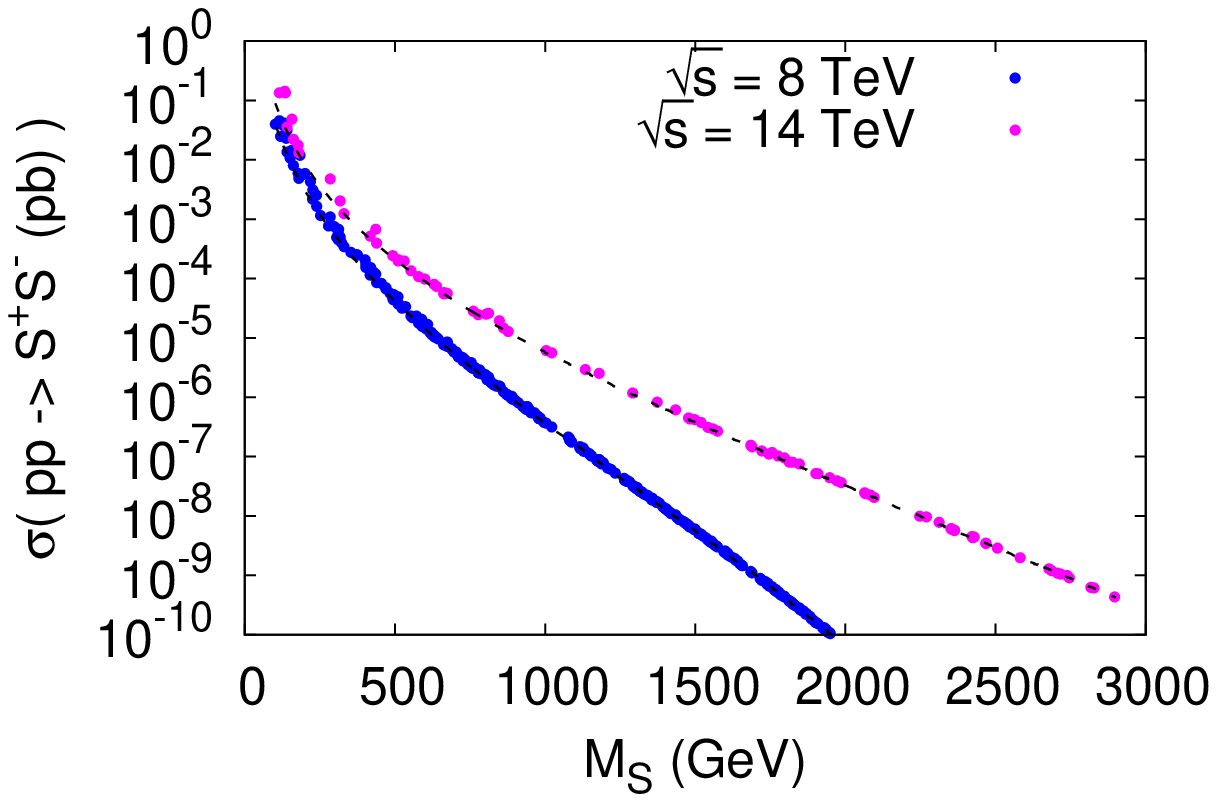}~\includegraphics[width = 0.5\textwidth,height=4.5cm]{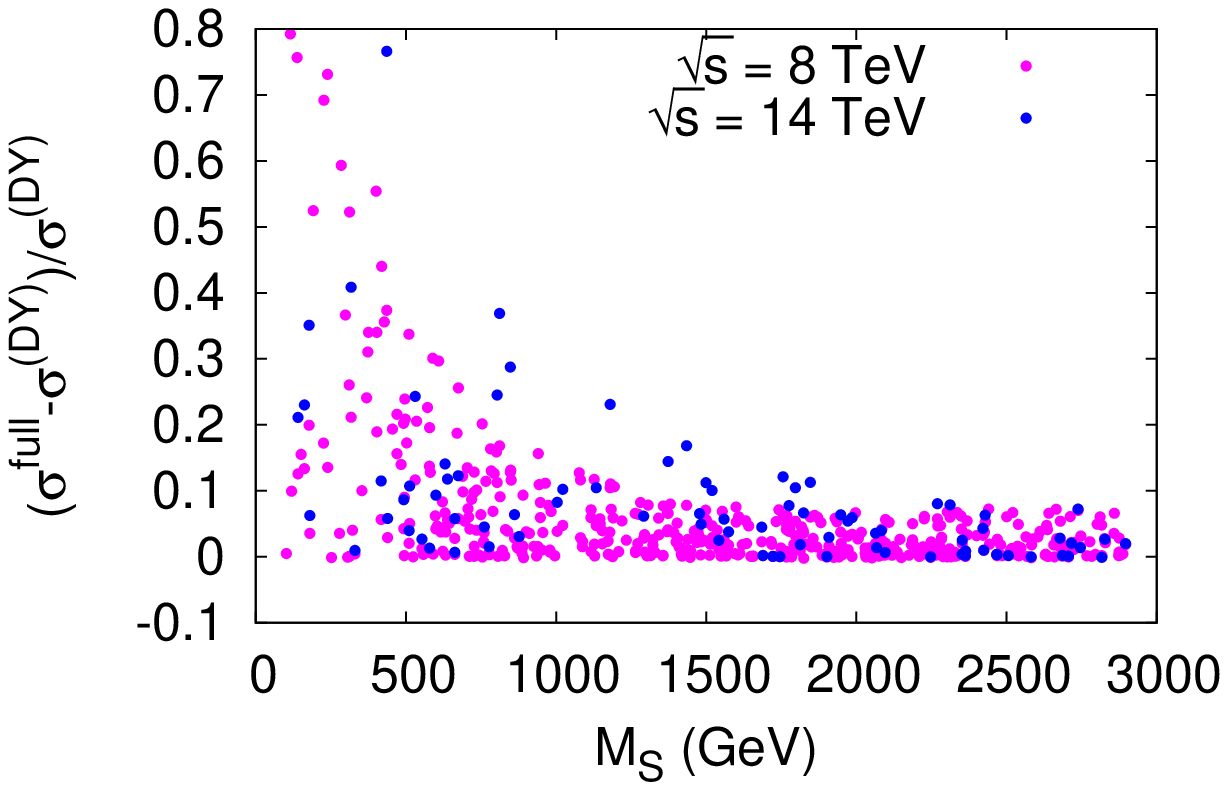}
\par\end{centering}
\caption{Left: the production cross section (in fb units) of $S^{+}S^{-}$ pair
production at $\sqrt{s}$ = 8 TeV (in magenta) and 14 TeV (in blue) vs the
charged scalar mass $M_{S}$. The dashed black lines represent the DY values.
Right: the ratio $\left[  \sigma^{full}(s)-\sigma^{(DY)}(s)\right]
/\sigma^{(DY)}(s)$ vs the charged scalar mass ($M_{S}$).}%
\label{knt-p}%
\end{figure}

The $f_{\alpha\beta}$ and $M_{S}$ model parameter are scanned randomly to explore the behaviour of the cross section of $S^{\pm}S^{\mp}$ pair production versus the charged scalar mass ($M_{S}$) at both $\sqrt{s}=8$ and $14~\mathrm{TeV}$ LHC energies in  Fig.~\ref{knt-p}. Furthermore, to estimate the dominant DY type diagrams we simply evaluate the cross sections ratio that characterises the presence of the Higgs Mediated Feynman diagrams ($\sigma^{(Full)}(s)\equiv\sigma(pp\rightarrow S^{+}S^{-})$ including the Higgs exchange diagrams).

\subsection{The charged scalar $S^{\pm}$ decay}

After its production, the charged scalar decays into a neutrino (manifested as missing energy in the detector) and a charged lepton (detected by the relevant sub-detector) with the partial decay rate ($\Gamma \varpropto \frac{|f_{\alpha\beta}|^{2}}{4\pi}~M_{S}$). At the LHC colliders, one can observe only three distinct signals since neutrinos are indistinguishable, i.e., charged lepton(s) and missing energy.
 
\begin{figure}[h]
\begin{centering}
\includegraphics[width=0.5\textwidth,height=4.5cm]{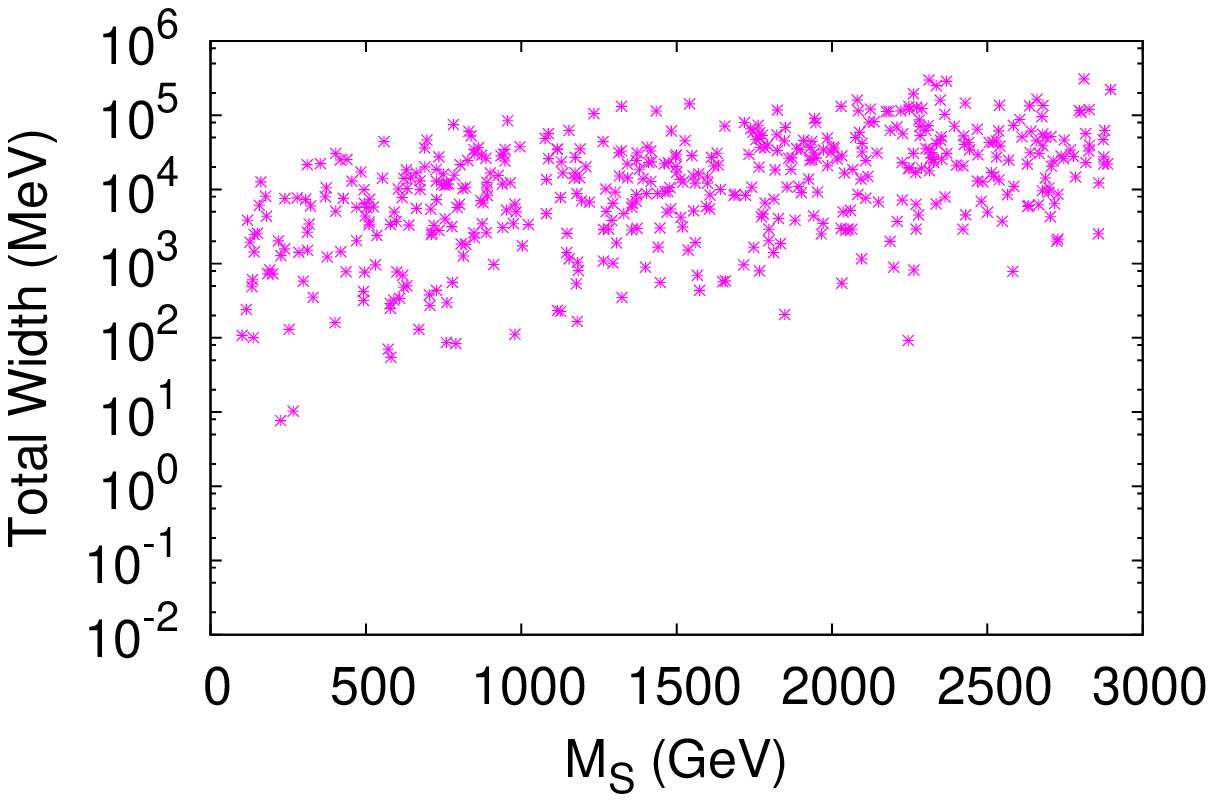}~\includegraphics[width = 0.5\textwidth,height=4.5cm]{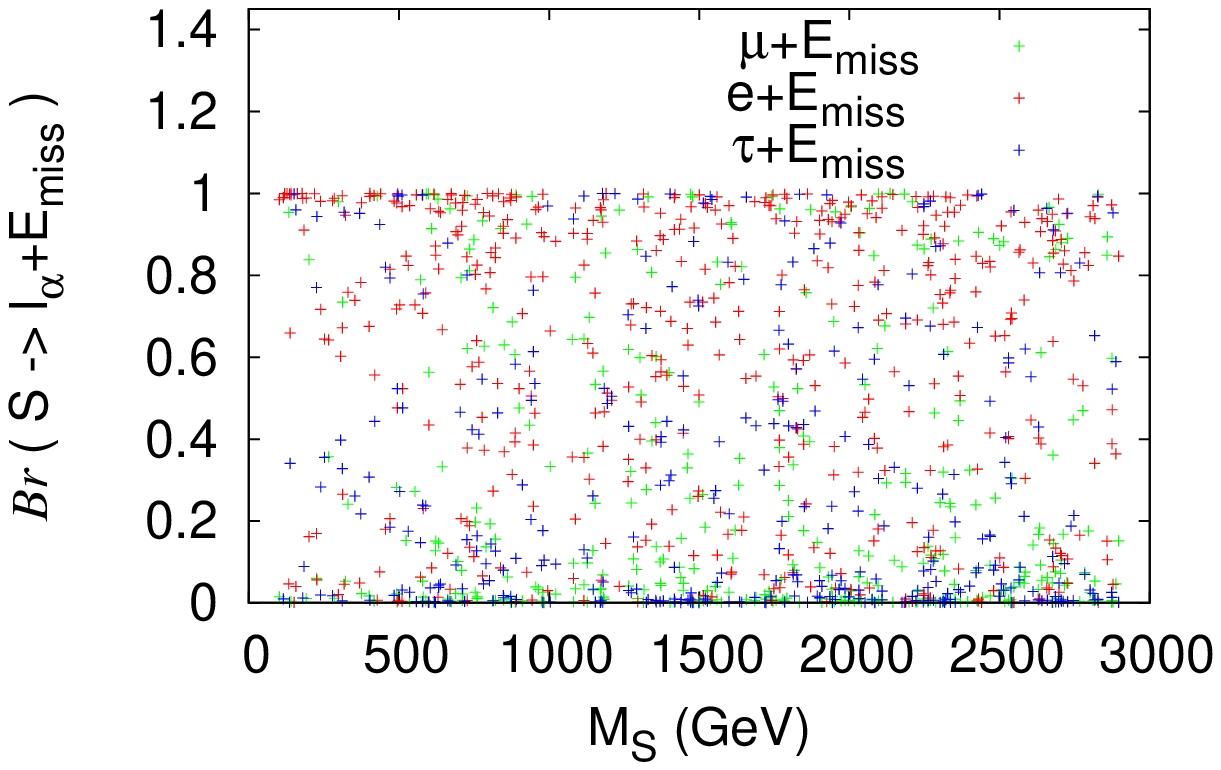}
\par\end{centering}
\caption{The charged scalar total decay width (left) and its different
branching ratios (right) vs $M_{S}$. The considered values of the charged
scalar mass and the Yukawa couplings $f_{\alpha\beta}$ are in agreement with LFV constraints.}%
\label{knt-d}%
\end{figure}

Considering similar benchmark points used in Fig.~\ref{knt-p}, we present in Fig.~\ref{knt-d} the total decay width of $S^{\pm}$ (left) and its branching ratios (right) for the different decay channels versus $M_{S}$.

\section{The charged scalar signatures at the LHC}
In this section, the possibility of testing this class of neutrino mass models (\ref{L}) via the LHC p-p collisions is studied. We will consider one benchmark given by the following parameter values%

\begin{equation}
\left\{
\begin{array}
[c]{l}%
f_{e\mu}=-(4.97+i1.41)\times10^{-2},\quad f_{e\tau}=0.106+i0.0859,\\
f_{\mu\tau}=(3.04-i4.72)\times10^{-6},\quad\quad M_{S}=914.2\textrm {GeV.}%
\end{array}
\right.  \label{fm}%
\end{equation}

The charged scalar signature at the LHC will be distinguished throughout the detection of two charged leptons plus missing energy ($pp\rightarrow\ell_{\alpha}^{\pm}\ell_{\beta}^{\mp}+ \slashed{E}$) where $\ell_{\alpha}^{\pm}\ell_{\beta}^{\mp}$ =$\{e^{+}e^{-}, e^{-}\mu^{+}, \mu^{-}\mu^{+}\}$. $\slashed{E}$ corresponds to any two SM neutrinos $\nu_{\alpha}\overline{\nu}_{\beta}$ ($\alpha, \beta = e,\mu, \tau$).
The main background comes from any process with the same previous final states where $WW$, $ZZ$, or $Z\gamma$ are intermediate states. The model files were built using LanHEP and then the event generation of both signal and background processes were simulated for various c.m. energies using CalcHEP.

\subsection{Signal Vs Background}

Any deviation from the SM in terms of cross section values will imply a significant detection of the $S^{\pm}$ signatures. In our case the compulsory cross section is the difference between the cross section evaluated in Eq. (\ref{L}) and the one evaluated within the SM framework. Consequently, an initial \textit{preselection} is performed in which a cut on the $M_{T2}$ \cite{mt2} variable is applied to eliminate the background charged leptons and neutrinos events coming from $W^{\pm}W^{\mp}$ ($M_{T2} > M_{W}$). The $M_{T2}$ is considered in the limit of massless missing energy particles as in Ref. \cite{Chatrchyan:2012jx} :

\begin{equation}
M_{T2}^{2}=2p_{T}^{\ell_{\alpha}}p_{T}^{\ell_{\beta}}\left(  1+\cos
\theta_{\alpha\beta}\right)  ,
\end{equation}

We study different kinematic distributions of the considered processes after imposing the $M_{T2}$ cut at $\sqrt{s}$ = 8 and 14 TeV, and then deduce the relevant cuts set as summarised in Tab ~\ref{Tab}.

\begin{table}[h]
\centering
\resizebox{\textwidth}{!}{%
\begin{tabular}
[c]{cclc}\hline
Process & Cuts@8 TeV &  & Cuts@14 TeV\\\hline\hline
$p$$p$ $\rightarrow$ $e^{-}$$\mu^{+}+ \slashed{E}$ & $%
\begin{array}
[c]{cc}%
80<p_{T}^{e^{-}}<250 & 80<p_{T}^{\mu^{+}}<270\\
-1.560<\eta_{e^{-}}<2.99 & -1.92<\eta_{\mu^{+}}<3
\end{array}
$ &  & $%
\begin{array}
[c]{cc}%
p_{T}^{e^{-}}>180 & p_{T}^{\mu^{+}}>170\\
1.1<\eta_{e^{-}}<2.89 & 1.2<\eta_{\mu^{+}}<3.02
\end{array}
$\\\hline\hline
$p$$p$ $\rightarrow$ $e^{-}$$e^{+}+ \slashed{E}$ & $%
\begin{array}
[c]{c}%
25<p_{T}^{l}<120\\
-2.09<\eta_{l}<2.89
\end{array}
$ &  & $%
\begin{array}
[c]{c}%
30<p_{T}^{l}<80\\
-2.8<\eta_{l}<2.95
\end{array}
$\\\hline\hline
$p$$p$ $\rightarrow$ $\mu^{-}$$\mu^{+}+ \slashed{E}$ & $%
\begin{array}
[c]{c}%
30<p_{T}^{l}<155\\
-2.38<\eta_{l}<2.1
\end{array}
$ &  & $%
\begin{array}
[c]{c}%
25<p_{T}^{l}<40\\
-0.13<\eta_{l}<3
\end{array}
$\\\hline
\end{tabular}
}
\caption{The considered cuts for the three final states at
$\sqrt{s}$= 8 and 14 TeV. The $p_{T}^{\ell}$
and $\eta_{\ell} $ are, respectively, the transverse momentum and
pseudorapidity of the charged lepton ($e,\mu$).} \label{Tab}
\end{table}

\subsection{Numerical results}

The separation of signal to background ratio for each final state using the selected cuts in Tab. \ref{Tab} is studied. The signal significance for each considered final state is given by

\begin{equation}
\begin{matrix}
  S=\frac{N_{ex}}{\sqrt{N_{ex}+N_{B}}} & \text{and} & N_{ex}=N_{M}-N_{B}=L\times(\sigma^{M}-\sigma^{B}),
\end{matrix} 
\end{equation}

where $N_{ex}$ denotes the excess events number of the considered signal, $N_{B}$ is the number of events of the background contributions and $N_{M}$ is the expected events number due to all the new model interactions including the SM contributions as well. $L$ indicates the integrated luminosity, and $\sigma^{M}$ ($\sigma^{B}$) is the total expected (background) cross section.
 
\begin{table}[h]
\centering%
\resizebox{\textwidth}{!}{%
\begin{tabular}
[c]{|c|cccc|c|cccc|}\hline
&  &  & $\sqrt{s}=8~TeV$ &  &  &  &  & $\sqrt{s}=14~TeV$ & \\\cline{2-5}%
\cline{7-10}%
Process & $\sigma^{M}(fb)$ & \multicolumn{1}{|c}{$\sigma^{B}(fb)$} &
\multicolumn{1}{|c}{$(\sigma^{M}-\sigma^{B})/\sigma^{B}$} &
\multicolumn{1}{|c|}{$S_{20}$} &  & $\sigma^{M}(fb)$ &
\multicolumn{1}{|c}{$\sigma^{B}(fb)$} & \multicolumn{1}{|c}{$(\sigma
^{M}-\sigma^{B})/\sigma^{B}$} & \multicolumn{1}{|c|}{$S_{100}$}\\\hline\hline
$pp\rightarrow e^{-}\mu^{+}+ \slashed{E}$ & 13.03 & \multicolumn{1}{|c}{11.98} &
\multicolumn{1}{|c}{0.0876} & \multicolumn{1}{|c|}{1.301} &  & 1.253 &
\multicolumn{1}{|c}{0.459} & \multicolumn{1}{|c|}{1.7} &
\multicolumn{1}{|c|}{7.093}\\\cline{1-5}\cline{7-10}%
$pp\rightarrow e^{-}e^{+}+ \slashed{E}$ & 62.74 & \multicolumn{1}{|c}{59.72} &
\multicolumn{1}{|c}{0.0506} & \multicolumn{1}{|c|}{1.7051} &  & 44.45 &
\multicolumn{1}{|c}{38.65} & \multicolumn{1}{|c|}{0.150} &
\multicolumn{1}{|c|}{8.699}\\\cline{1-5}\cline{7-10}%
$pp\rightarrow\mu^{-}\mu^{+}+ \slashed{E}$ & 81.691 & \multicolumn{1}{|c}{77.49}
& \multicolumn{1}{|c}{0.0542} & \multicolumn{1}{|c|}{2.0786} &  & 65.27 &
\multicolumn{1}{|c}{56.86} & \multicolumn{1}{|c|}{0.148} &
\multicolumn{1}{|c|}{10.409}\\\hline
\end{tabular}
}
\caption{The cross section of the total expected signals
($\sigma^{M} $) and the corresponding background ($\sigma^{B}$) are
used to estimate the significance $S_{20}$ ($S_{100}$)\ at 8
\textrm{TeV} (14 \textrm{TeV}) with $L=20~fb^{-1}$ ($L=100~fb^{-1}$).}
\label{tab-S}%
\end{table}

\begin{figure}[h]
\begin{center}
\includegraphics[width=0.46\textwidth,height=4cm]{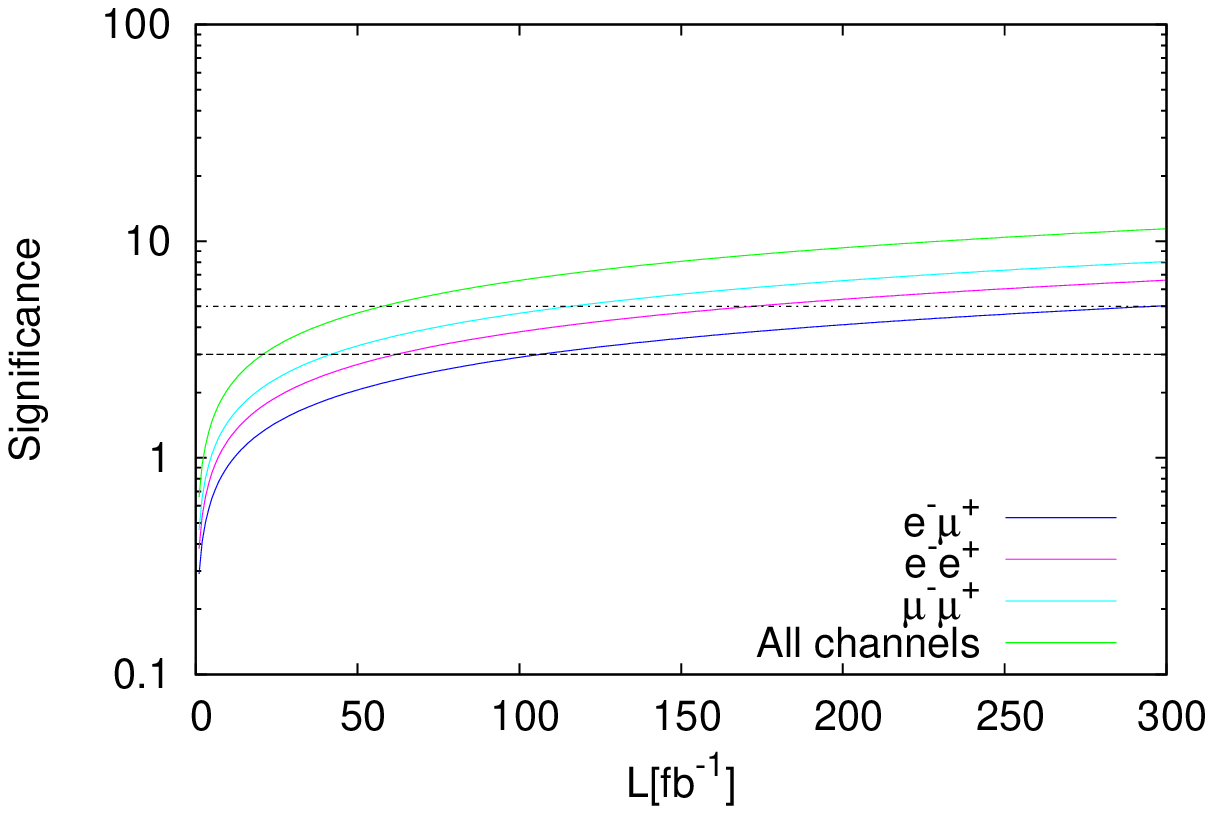}~\includegraphics[width=0.46\textwidth,height=4cm]{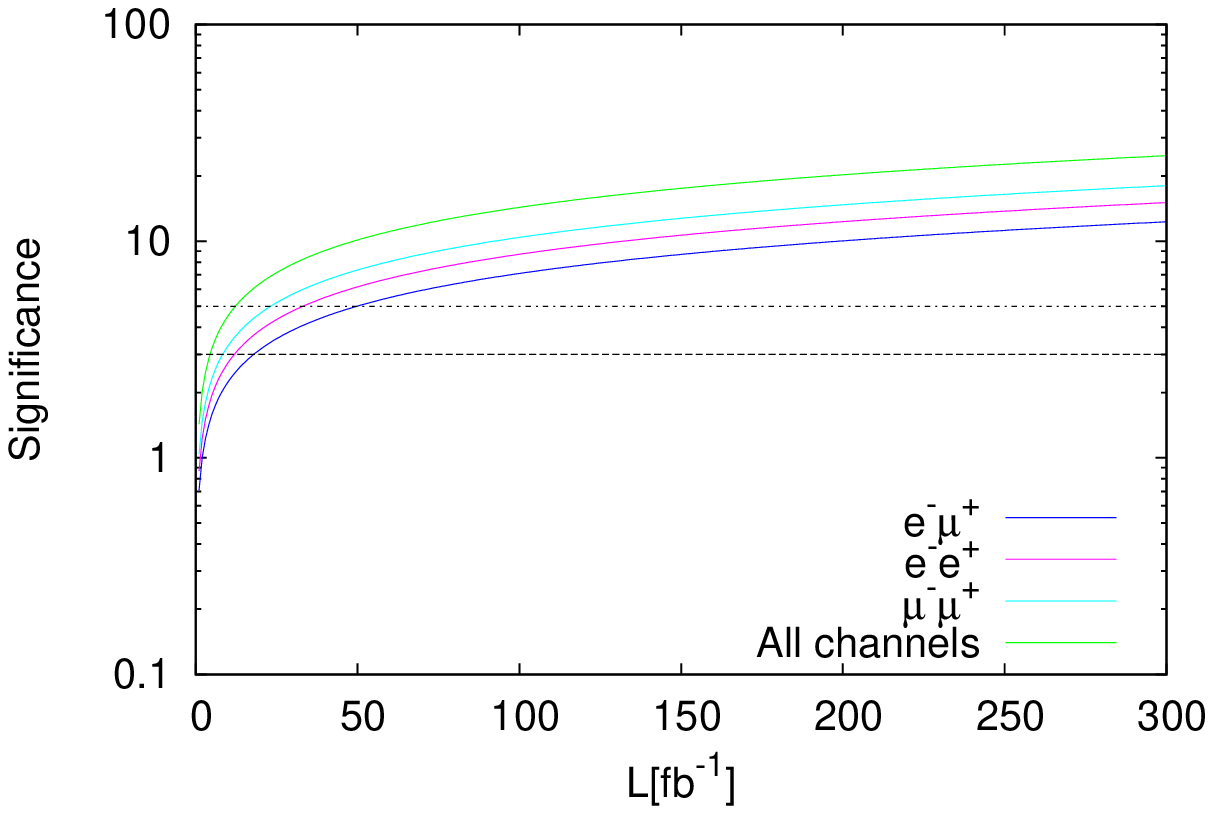}
\end{center}
\caption{The significance vs the luminosity at $\sqrt{s}=8~TeV$ (left) and
at $\sqrt{s}=14~TeV$ (right) for each signature. The two horizontal lines in each panel indicate the corresponding significance values for $S=3$ and $S=5$.}%
\label{l}%
\end{figure}

\begin{figure}[h]
\begin{center}
\includegraphics[width=0.46\textwidth,height=4cm]{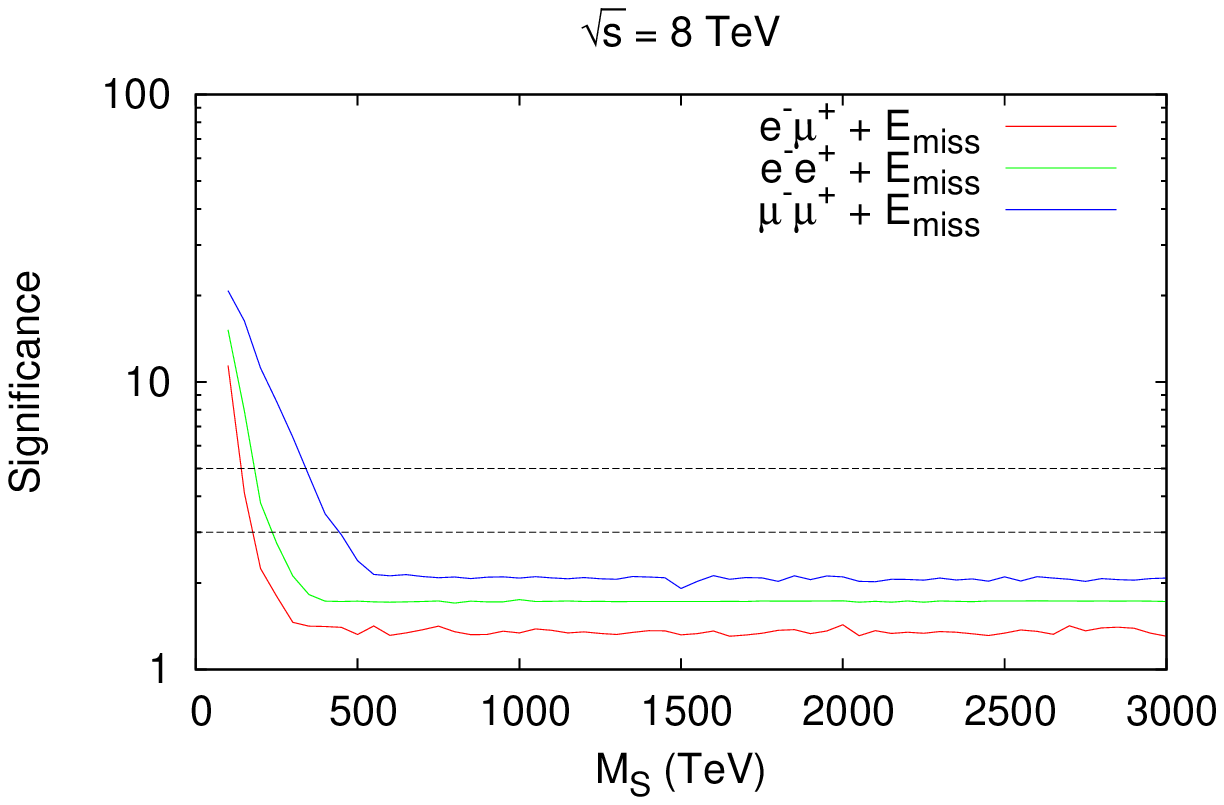}~\includegraphics[width=0.46\textwidth,height=4cm]{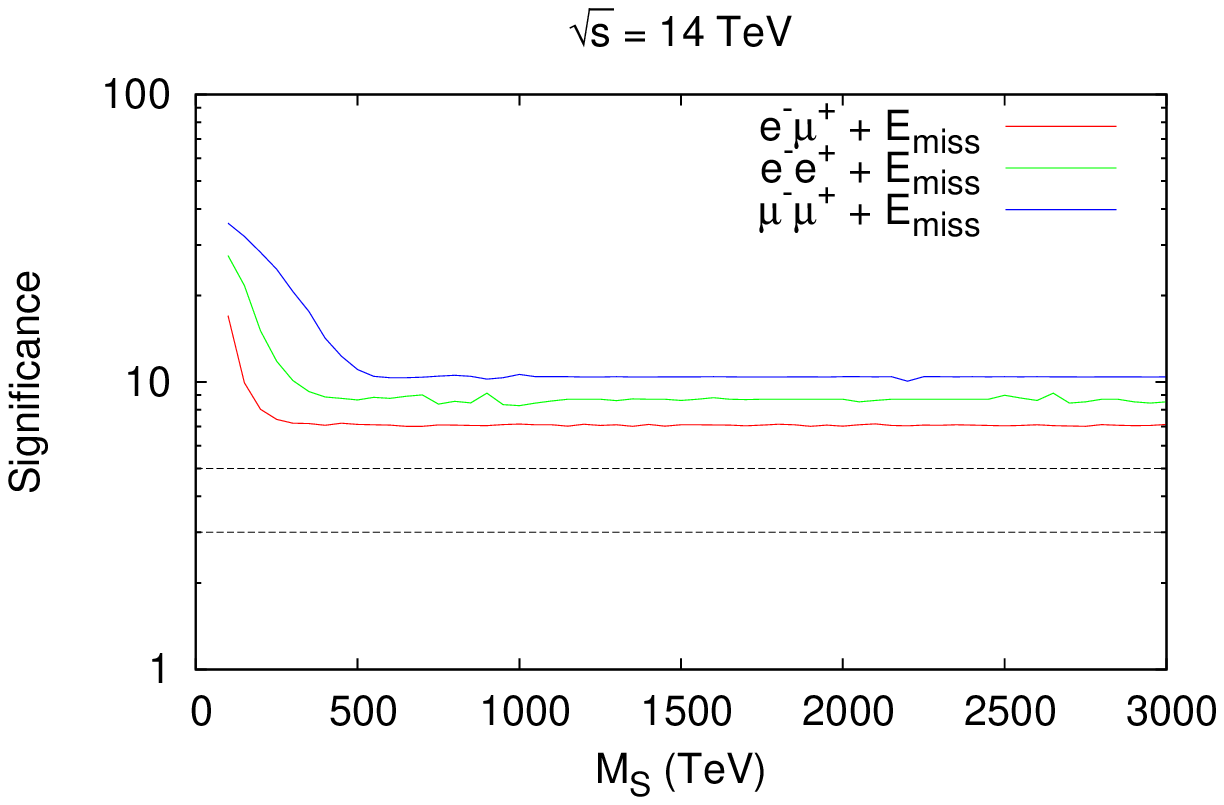}
\end{center}
\caption{The significance vs $M_{S}$ for each channel at L = 20 $fb^{-1}$ (left) and at L = 100 $fb^{-1}$ (right). The two horizontal lines indicate the
significance values at S = 3 and S = 5, and the vertical corresponds to $M_{S}$=914 GeV.}
\label{ms1}
\end{figure}

After imposing the cuts set given above, the results in Tab ~\ref{tab-S} show the variation of the signal significance within the range of [1.30 - 2.07] at both energies $\sqrt{s}=$ 8 and 14 \textrm{TeV}. These results can be also illustrated as in Fig.~\ref{l} where the significance is plotted versus the luminosity. 
One direct hint from Fig.~\ref{l}, is to look at the $S^{\pm}S^{\mp}$ production in the $\mu^{-}\mu^{+}$ channel that is significantly larger than $e^{-}e^{+}$ channel. Furthermore, one can study the dependence of the significance versus $M_{S}$ with respect LFV bounds at $\sqrt{s}$ = 8 and 14 TeV as shown in Fig.~\ref{ms1}. Hence, in Fig.~\ref{ms1}, left, the LHC Run I data can be used to exclude $M_{S}$ $<$ 400 GeV, whereas Fig.~\ref{ms1}, right, shows the same lower bound on $M_{S}$ with more than 5 $\sigma$ significance.

\section*{Acknowledgments}
C. Guella would like to thank the organisers of the 51th Rencontres de Moriond EW interactions and unified theories for the inspiring atmosphere during the workshop and for the full support.

\end{document}